\documentstyle[prl,multicol,aps]{revtex}
\begin{document}

\draft
\title{ A fourfold coordinated point defect in silicon }
\author{Stefan Goedecker, Luc Billard, Thierry Deutsch}
\address{ D\'epartement de recherche fondamentale sur la mati\`ere condens\'ee,\\
         SP2M/NM, CEA-Grenoble, 38054 Grenoble cedex~9, France }
\date{\today}
\maketitle

\begin{abstract}
Due to their technological importance, point defects in silicon 
are among the best studied physical systems. 
The experimental examination of point defects buried in bulk is difficult 
and evidence for the various defects usually indirect. Simulations of 
defects in silicon have been performed at various levels of sophistication 
ranging from fast force fields to accurate 
density functional calculations. The generally accepted viewpoint from all 
these studies is that vacancies and self interstitials are the basic 
point defects in silicon. We challenge this point of view by 
presenting density functional calculations that show that there is a new 
fourfold coordinated point defect in silicon that is lower in energy. 
\end{abstract}

\begin{multicols}{2}
\setcounter{collectmore}{5}
\raggedcolumns

\begin{figure}[h]             
     \begin{center}
      \setlength{\unitlength}{1cm}
       \begin{picture}( 10.0,16.0)           
        \put(-2.5,8.5){\includegraphics{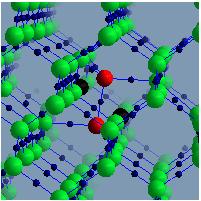}}   
        \put(0,8.4){FFCD defect}   
        \put( 5.5,8.5){\includegraphics{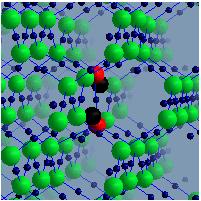}}   
        \put( 8,8.4){FFCD defect}
        \put(-2.5,0.5){\includegraphics{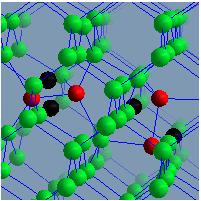}}   
        \put(-1,0.4){Two non-parallel FFCD's }  
        \put( 5.5,0.5){\includegraphics{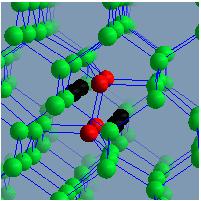}}   
        \put( 7,0.4){Two parallel FFCD's }
       \end{picture}
       \caption{Two views of the FFCD defect (upper part) and two FFCD pairs 
                (lower part). The defects are obtained by moving 
                atoms from the initial positions denoted by black spheres 
                into the final red positions. The centers of the Wannier functions 
                are indicated by the small blue spheres for the single FFCD but 
                omitted for the two pairs. 
                The formation of the FFCD is easily understandable from the upper 
                left panel. There, 
                the upper red atom originates form the upper black position to the left. 
                The swinging movement needed for this displacement does not 
                break the bonds with the upper neighbors, but with the lower ones.
                A corresponding swing is needed for the lower defect atom.
                In the final position two new bonds can be formed  
                so that in the end all atoms are again fourfold coordinated. 
                              \label{const}}
      \end{center}
     \end{figure}

The stability of crystalline silicon comes from the fact that each silicon atom can 
accommodate its four valence electrons in four covalent bonds with its four neighbors. 
The traditional point defects in silicon, the vacancy and the various interstitials, 
are obtained by taking out or adding atoms to the crystal and thus all destroy 
fourfold coordination. A relatively high defect formation energy for these defects 
is the consequence. In addition there is a point defect, that conserves the 
number of particles, the Frenkel pair, consisting of a vacancy and an interstitial.
If the vacancy and the interstitial are close, the formation energy of a Frenkel pair 
is less than the sum of an isolated vacancy and interstitial (Table~\ref{energies}). 
Nevertheless the formation energy is still considerable since again bonds are broken. 
Figure~\ref{const} shows a novel defect configuration that has, in contrast to 
all other point defects, perfect fourfold 
coordination and will therefore be called Four Fold Coordinated Defect (FFCD).

Using state of the art plane wave density functional theory (DFT) calculations 
we will now present evidence that the formation energy of the FFCD 
of 2.4 eV is lower than 
the formation energy of all other known point defects both in intrinsic and 
doped silicon.  Even though several calculations of this type were published for the 
traditional defects, we have decided 
to repeat them for several reasons. a) All these calculations take advantage 
of error cancellations to obtain energy differences that are more precise than 
the total energies themselves. This cancellation is obviously best if all 
the total energies that are compared are calculated with exactly the same method. 
b) Most DFT calculations 
used the basic Local Density Approximation (LDA) whereas we used a more precise 
General Gradient Approximations~\cite{pbe}. c) As was recently shown~\cite{puska} cells 
of at least 216 atoms are required for a reasonable convergence. Most 
published calculations were done with smaller cells. d) We constructed very accurate 
pseudopotentials~\cite{psp} based on atomic calculations with the same density 
functional as used in our target calculation and we afforded a 
very large plane wave basis set (25 Ry for the LDA and 35 Ry for the GGA calculations). 
e) The point defect formation energies in doped silicon were up to now obtained  
in an approximate way by combining density functional total energy results 
with the concept of a chemical potential that varies as a function of doping 
across the experimental band gap of 1.12 eV of 
silicon~\cite{car,lee}. We have modeled the effect 
of doping in a coherent way for both n and p type doped silicon by including 
explicitly two doping atoms (aluminum and phosphor) in our 216 atom supercell. 
This procedure leads of course to high doping concentrations for 
any reasonable size of our computational cell, but by treating coherently 
and accurately the extreme cases of high and zero doping 
we can expect the numbers for more moderate doping concentrations to 
lie in between these results.

Table~\ref{energies} shows all the total energy results. 
We have included in our compilation all defects that have been classified as 
being low in energy in previous calculations~\cite{lee,car,bloechl,kaxiras}, 
i.e. the split (110) X interstitial, the hexagonal H interstitial 
and the vacancy. In addition we have included a low energy Frenkel pair. 
In agreement with recent calculation by Needs~\cite{needs,qmc}, we found that both 
the tetrahedral interstitial and the caged interstitial~\cite{ackland} are higher in 
energy and meta-stable. They are therefore not included in the compilation.
The energies in Table~\ref{energies} do not contain an additional lowering of the 
energy that is obtained in an infinite crystal by a long range elastic field, 
which is suppressed in our finite cell. We have calculated this extra elastic 
energy for a few configurations and found it to be negligible, namely less than .05 eV. 
Because of the negative U character of the traditional vacancies and interstitials 
that was postulated with the  help of 
DFT calculations~\cite{baraff,joann} 
and confirmed experimentally~\cite{negu},  
the lowest energies are obtained by closed shell configurations.
Consequently all calculations were performed without spin polarization.

For the FFCD the bond length and angle do not 
significantly deviate from their bulk values. 
Whereas the bond length and angle in the bulk are 2.35 \AA $\:$ and 109 degrees, they 
respectively vary from 2.25 to 2.47 \AA $\:$ and from 97 to 116 degrees for the bonds  
formed by the two (red) defect atoms. The fourfold coordination is also 
visible from the fact that the centers of the maximally localized Wannier 
functions~\cite{marzari}, that give the best possible spatial location of 
an electron,  are located practically exactly in the middle of 
the bonds established by the geometric distance criteria. The spread of 
all the Wannier functions is very close to the bulk value of 2.7 a.u. 
The fact that the LDA and GGA values are very close is not surprising.
The simple covalent bonds are very well described by any reasonable 
density functional and we expect an error of less than a tenth of an eV 
for this configuration.

Correlation effects are more important for the broken bonds of the other point 
defects. Hence the DFT values are less reliable than in the case of the FFCD. 
Recent Quantum Monte Carlo(QMC) calculations~\cite{qmc} for interstitials are in much 
better agreement with GGA values than with LDA values. 
We therefore believe that it is unlikely that the energetical ordering 
predicted by our GGA calculations does not correspond to reality.

The broken bonds of the other point defects are again best visualized by looking at the 
centers of the maximally localized Wannier functions in the neighborhood of 
defects. They are located in empty space regions rather than at positions 
where geometric criteria would predict bonds. Their spread is also significantly 
larger than the bulk value. 
In the case of the vacancy the two non-bonding Wannier functions have a 
spread of 3.6. a.u., for the X interstitial 4.1 a.u. and for the H 
interstitial 4.7 a.u.. For the Frenkel pair there are two 
non-symmetric Wannier functions with spreads of 3.5 and 4.4 a.u..

Whereas the classical point defects were all postulated based on simple 
symmetry considerations, the discovery of this new point defect was 
made possible by new algorithmic developments. 
The configurational space was systematically explored by using a modified 
basin hopping method~\cite{hopp} and an inter-atomic silicon potential~\cite{lenos}.
The most promising configurations were then refined with a plane wave 
electronic structure code~\cite{cpmd}  
developed by J. Hutter et al.. The preconditioned 
conjugate gradient method~\cite{payne} was used for the electronic optimization 
and the DIIS method~\cite{pulay} without any symmetry constraints 
for the ionic relaxation. 

\begin{figure}[h]             
     \begin{center}
      \setlength{\unitlength}{1cm}
       \begin{picture}( 8.0,12.0)           
        \put(-3.,0.0){\includegraphics{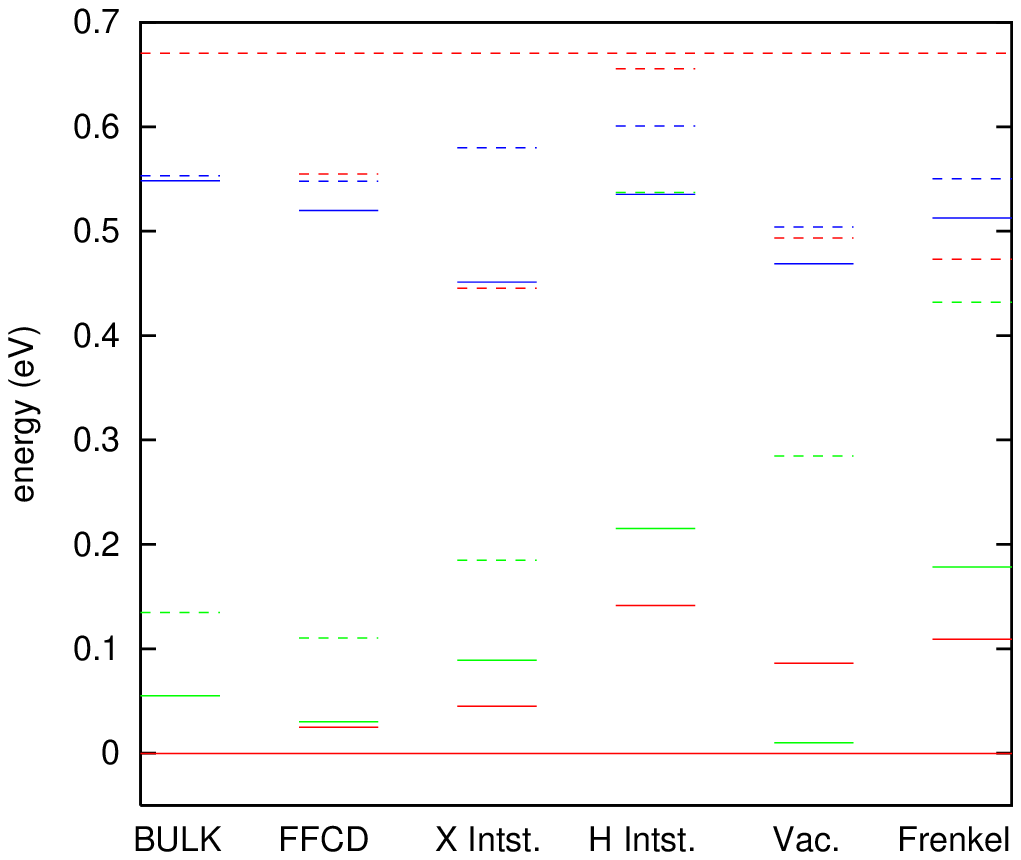}}   
       \end{picture}
       \caption{The HOMO LUMO splitting for the configurations of Table~\ref{energies}.
                The HOMO is denoted by a solid line, the LUMO by a 
                dashed line. Red corresponds to intrinsic silicon, 
                green to p-doped and blue to n-doped silicon. The two 
                levels of intrinsic silicon without defects are drawn over the 
                whole interval. \label{gap}}
      \end{center}
     \end{figure}

The remarkable fact that the FFCD has not yet been detected experimentally, in spite 
of its equilibrium concentration which should be many orders 
of magnitude larger than that of the other defects, is probably due to 
two factors. First, the experimental search for defects was always guided 
by theoretical predictions. Second, it is invisible with 
standard experimental techniques.  Because of the 
perfect fourfold coordination no unpaired electrons exist, a necessary 
condition for Electronic Paramagnetic Resonance (EPR) and  
Electron Nuclear DOuble Resonance (ENDOR) experiments.  Deep Level Transient 
Spectroscopy (DLTS) requires electronic levels within the gap. Given the perfect 
coordination this seems unlikely. To examine this point further we have calculated 
the LDA Highest Occupied Molecular Orbital (HOMO),
Lowest Unoccupied Molecular Orbital (LUMO) splitting.
Even though it is known that 
the gap is not well reproduced by DFT, these levels allow for a qualitative 
interpretation.
The results (Figure~\ref{gap}) show that the FFCD disturbs the band structure 
indeed much less than the other defects. 
The FFCD spectrum is also insensitive to the doping level whereas the other spectra 
are strongly influenced via various Jahn Teller distortions~\cite{bachelet} because 
of their orbital degeneracy. 

The basic FFCD can serve as a building block for more extended defects. 
Two simple defects obtained by combining just two FFCD's in two different ways are 
shown at the bottom of Figure~\ref{const}. Their energies of 3.98 eV for the 
tilted configuration 
and of 4.17 eV for the parallel configuration are significantly lower than the sum of 
the energies of two isolated FFCD's.

The consequences of our results are wide ranging. A reexamination of 
numerous experimental results will be necessary. In particular the tacit 
assumption that the same kind of defects are responsible both for the 
electrical and diffusion properties of silicon has been refuted by our 
calculations. Even though the FFCD  does presumably not strongly influence 
transport properties, it is expected to play an important role in the 
diffusion properties of silicon.

We acknowledge the most useful comments of Giovanni Bachelet, 
Erik Koch, Richard Needs and Cyrus Umrigar on our manuscript.

\end{multicols}{2}

\pagebreak
\begin{table}[h]
\caption[]{ The GGA formation energy in eV of the point defects  
described in the text and depicted in Figures~\ref{const} and ~\ref{nconst} 
For comparison the LDA values are given in parentheses for intrinsic silicon. 
\label{energies}}
\begin{tabular}{|l||c|c|c|} \hline
                &   p-type    &  intrinsic  & n-type      \\ \hline
 FFCD           &        2.45 & 2.42 (2.34) &  2.39 \\ \hline
 Frenkel        &        5.65 & 4.32 (4.26) &  5.77 \\ \hline
 X interstitial &        3.33 & 3.31 (2.88) &  2.98 \\ \hline
 H interstitial &        2.80 & 3.31 (2.87) &  3.12 \\ \hline
 Vacancy        &        3.01 & 3.17 (3.56) &  3.14 \\ \hline
\end{tabular}
\end{table}

\pagebreak

   \begin{figure}[h]             
     \begin{center}
      \setlength{\unitlength}{1cm}
       \begin{picture}( 10.0,16.0)           
        \put(-2.5,8.5){\includegraphics{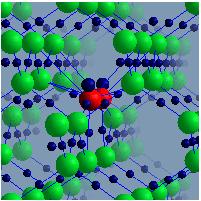}}   
        \put(0,8.4){X interstitial}   
        \put( 5.5,8.5){\includegraphics{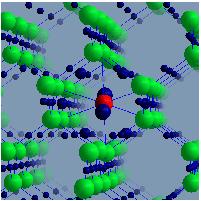}}   
        \put( 8,8.4){H interstitial}
        \put(-2.5,0.5){\includegraphics{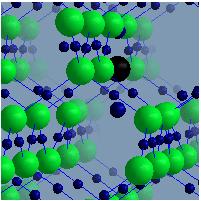}}   
        \put(0,0.4){Vacancy}  
        \put( 5.5,0.5){\includegraphics{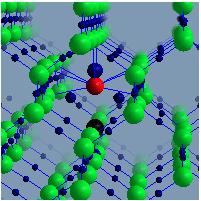}}   
        \put( 8.0,0.4){Frenkel pair}
       \end{picture}
       \caption{Classical point defects in silicon: 
                Red atoms indicate atoms in new positions.
                For the vacancy and the Frenkel pair the empty lattice 
                site is shown by a black sphere.
                The centers of extended Wannier functions are 
                visualized by increasingly bigger blue 
                spheres, Wannier functions with the typical bulk extension 
                by small blue spheres.\label{nconst}}
      \end{center}
     \end{figure}

\begin{thebibliography}{}

\bibitem{pbe}  Perdew, J., Burke, K. \& Ernzerhof, M.,
Generalized Gradient Approximation Made Simple, 
Phys. Rev. Lett. {\bf 77} 3865 (1996).

\bibitem{puska}
Puska, M., P\"{o}ykk\"{o}, S., Pesola, M. \& Nieminen, R., 
Convergence of supercell calculations for point defects in semiconductors: 
Vacancy in silicon, 
Phys. Rev. B {\bf 58} 1318 (1998)

\bibitem{psp}  
Goedecker, S., Teter, M. \& Hutter, J., 
Separable dual space Gaussian Pseudopotentials,
Phys. Rev. B {\bf 54}, 1703 (1996)

\bibitem{lee}
Lee, W., Lee, S. \& Chang, K., 
First-principles study of the self-interstitial diffusion mechanisms in silicon, 
J. Phys.: Condens. Matter {\bf 10} 995 (1998)

\bibitem{car}
Car, R., Kelley, P., Oshiyama, A.,  \&  Pantelides, S.,  
Microscopic theory of Impurity-defect reactions and impurity diffusion in silicon, 
Phys. Rev. Lett. {\bf 52} 1814 (1984)

\bibitem{bloechl}
P. Bl\"{o}chl, E. Smagiassi, R. Car, D. Laks, W. Andreoni \& S. Pantelides,
First-principles calculations of self-diffusion constants in silicon, 
Phys. Rev. Lett. {\bf 70} 2435 (1993)

\bibitem{kaxiras}
Antonelli, A., Kaxiras, E. \& Chadi, D., 
Vacancy in silicon revisited: Structure and Pressure effects, 
Phys. Rev. Let. {\bf 81} 2088 (1998)

\bibitem{needs}
Needs, R., 
First-principles calculations of self-interstitial defect structures and diffusion 
paths in silicon, 
J. Phys.: Condens. Matter {\bf 11} 10437 (1999)

\bibitem{qmc}
Leung, W., Needs, R., Rajagopal, G., Itoh, S. \&  Ihara, S., 
Calculations of silicon self-interstitial defects, 
Phys. Rev. Lett. {\bf 83} 2351 (1999)

\bibitem{ackland}
S. Clark \& G. Ackland, 
Ab initio calculations of the self-interstitial in silicon, 
Phys. Rev. B {\bf 56} 47 (1997)

\bibitem{baraff}
Baraff, G.,  Kane, E., \& Schl\"{u}ter, M., 
Silicon vacancy: A possible Anderson negative U system, 
Phys. Rev. Lett. {\bf 43} 956 (1979)

\bibitem{joann}
Bar-Yam, Y. \& Joannopoulos, J., 
Silicon self-interstitial migration: Multiple paths and charge states, 
Phys. Rev. B {\bf 30} 2216 (1984)

\bibitem{negu}
Watkins, G., Troxell, J., 
Negative-U properties for point defects in silicon
Phys. Rev. Lett. {\bf 4} 593 (1980)

\bibitem{marzari}
Marzari, N. \& Vanderbilt, D., 
Maximally localized generalized Wannier functions for composite energy bands, 
Phys. Rev. B {\bf 56}, 12847 (1997)

\bibitem{bachelet}
Bachelet, G., in "{\it Crystalline Semiconductor Materials and Devices}", edited 
by P. Butcher and N. March, Plenum Publishing, New York, 1986

\bibitem{hopp}
Doye, J., \& Wales, D., Thermodynamics of global optimization, 
Phys. Rev. Lett. {\bf 80} 1357 (1998)

\bibitem{lenos}
Lenosky, T. et al., Highly optimized empirical model of silicon, 
Modelling Simul. Mater. Sci. Eng. {\bf 8}, 825, (2000)

\bibitem{cpmd} 
CPMD Version 3.3: developed by J. Hutter, A. Alavi, T. Deutsch, 
M. Bernasconi, S. Goedecker, D. Marx, M. Tuckerman and M. Parrinello, 
Max-Planck-Institut f\"{u}r Festk\"{o}rperforschung and IBM Z\"{u}rich Research Laboratory 
(1995-1999)

\bibitem{payne}
M. Payne, M. Teter, D. Allan, T. Arias \& J. Joannopoulos, 
Iterative minimization techniques for ab initio total-energy calculations: 
molecular dynamics and conjugate gradients, 
Rev. of Mod. Phys. {\bf 64}, 1045, (1992) 

\bibitem{pulay}
Pulay, P., Convergence acceleration of iterative sequences, 
Chem. Phys. Lett., {\bf 73}, 393, (1980)

\end{thebibliography}
\end{document}